# Commercial AI, Conflict, and Moral Responsibility

A theoretical analysis and practical approach to the moral responsibilities associated with dual-use AI technology

## Daniel Trusilo

University of California, San Diego, dtrusilo@ucsd.edu

## David Danks

University of California, San Diego, ddanks@ucsd.edu

This paper presents a theoretical analysis and practical approach to the moral responsibilities when developing AI systems for non-military applications that may nonetheless be used for conflict applications. We argue that AI represents a form of crossover technology that is different from previous historical examples of dual- or multi-use technology as it has a multiplicative effect across other technologies. As a result, existing analyses of ethical responsibilities around dual-use technologies do not necessarily work for AI systems. We instead argue that stakeholders involved in the AI system lifecycle are morally responsible for uses of their systems that are *reasonably foreseeable*. The core idea is that an agent's moral responsibility for some action is not necessarily determined by their intentions alone; we must also consider what the agent could reasonably have foreseen to be potential outcomes of their action, such as the potential use of a system in conflict even when it is not designed for that. In particular, we contend that it is reasonably foreseeable that: (1) civilian AI systems will be applied to active conflict, including conflict support activities, (2) the use of civilian AI systems in conflict will impact applications of the law of armed conflict, and (3) crossover AI technology will be applied to conflicts that fall short of armed conflict. Given these reasonably foreseeably outcomes, we present three technically feasible actions that developers of civilian AIs can take to potentially mitigate their moral responsibility: (a) establishing systematic approaches to multi-perspective capability testing, (b) integrating digital watermarking in model weight matrices, and (c) utilizing monitoring and reporting mechanisms for conflict-related AI applications.

## 1 INTRODUCTION

The increasing ubiquity of Artificial Intelligence (AI) systems, and the potential that they will be used or adapted for conflict operations, has been widely recognized. Historically, discussions of military AI systems have focused on autonomous weapons systems, but we are now seeing more recognition that there are other ways in which AI can impact conflicts of all types. For example, the White House Executive Order on *Safe, Secure, and Trustworthy Development and Use of AI* highlights not only risks for consumers and the public, but also the substantial potential dual-use (i.e., civilian and military) risks presented by foundation models [1]. Similarly, a document prepared by the government of the United Kingdom for the 2023 AI Safety Summit concludes that AI is a force multiplier, "proliferating and enhancing threat actor capabilities and increasing the speed, scale and sophistication of attacks" [2]. Despite the recognition of these issues, there has been little clarity on how to think about potential moral and practical responsibilities that arise from recognition of the multi-functionality of an AI system. If we acknowledge that AI systems can be dual-use (civilian and military), then what ought we do differently in design, development, deployment, monitoring, and so forth? We aim here to formulate a coherent theoretical and practical approach to the moral responsibilities when developing AI systems for non-military applications that may nonetheless be used for conflict applications due to their inherent dual-use nature. Because we are principally interested in ethical obligations and challenges that arise in the dual- or multi-use case, we exclude AI systems (and their developers) that are explicitly designed for security applications, including autonomous weapons systems.

Policy debates and discussions often differentiate between commercial AI for civilian purposes and systems designed for military applications. For example, the draft EU AI Act explicitly excludes applications designed for military use: "AI systems exclusively developed or used for military purposes should be excluded from the scope of this Regulation where that use falls under the exclusive remit of the Common Foreign and Security Policy regulated under Title V of the Treaty on the European Union" [3]. And the White House Executive Order on the Safe, Secure, and Trustworthy Development and Use of Artificial Intelligence alludes to special concerns about national security and dual-use technology, but not in terms of its application to armed conflict [1]. Although these different uses of AI technology are often treated separately, we argue in Section 2 that military uses of AI technologies cannot so easily be carved out. That is, current policies and governance documents depend on a blurry, and perhaps non-existent, boundary. Moreover, the challenge is even more complicated when we look beyond military use of civilian AI technology to commercial AI uses by *any* actor involved in conflict, including state proxy forces, non-state armed groups, mercenaries, quasi-security forces, and many others. Given the range of different conflict-related actors who might use a commercial AI system, we ask: **what are the ethical implications for, and obligations on, developers of commercial AI systems?**

We start in Section 2 by examining the novel threats that civilian AI systems can pose for conflicts, including qualitative differences with previous dual-use technologies. Section 3 then explores the resulting moral and ethical responsibilities of actors involved in the development of civilian AI systems (i.e., not intended for conflict uses) by extending the notion of *reasonably foreseeable outcomes* to the use of such AI systems. In Section 4, we argue that if particular risks or harms are reasonably foreseeable, then actors have (at least) three interconnected practical responsibilities: (a) establish systematic approaches to multi-stakeholder capability testing (similar to so-called AI red teaming), (b) integrate digital watermarking in model weight matrices, and (c) create and utilize monitoring and reporting mechanisms for conflict-related AI applications. We argue that these three sets of actions, when carried out with reasonable diligence, are appropriate for stakeholders to meet, or at least largely address, their moral obligations with regards to their technology. Moreover, these actions are all technically feasible, though they would have impacts across the full AI system lifecycle.

## 2 ESTABLISHING THE NOVEL ASPECTS OF AI AS A CROSSOVER TECHNOLOGY

Governments [2], academics [4], and private corporations [5] acknowledge that civilian AI technology, even when not specifically designed for security applications, can offer capabilities that could be beneficial to actors involved in conflict. For example, the Ukraine conflict has seen extensive use of commercial drones with AI-supported autonomous functions [6], Maxar's synthetic aperture radar technology for imagery [7], and Clearview AI's facial recognition software to identify combatants [8]. More generally, there is a growing body of research that shows that AI technologies not normally associated with military uses (i.e., AI besides autonomous targeting or weapons systems) nonetheless can offer powerful capabilities for conflict applications, partly because they are often relatively inexpensive, easy to adopt, and readily available [4, 9, 10]. The general idea of this type of *crossover AI technology*—i.e., technology that starts in the civilian domain and finds uses in conflict settings—is not new, as it is simply a generalization of the notion of dual-use technology. We contend, though, that crossover AI technology is different than previous historical examples of dual- or multi-use technology.

## 2.1 AI as a different type of crossover technology

There are a multitude of historical examples of commercial civilian technology crossing over into conflicts (e.g., fireworks, the combustion engine, flight technology, lasers, etc.). Crossover *AI* technologies are fundamentally different, however, because AI can be applied across other technologies: it is a multiplicative enhancer of existing capabilities, in addition to providing novel capabilities of its own. Eric Schmidt frames this novel impact of AI technology in terms of *innovation power* arguing that AI is contributing to an unprecedented speed of innovation and that "the ability to invent, adopt, and adapt new technologies," is the defining new force of international politics [11]. The former director of the US Department of Defense (DoD) Joint Artificial Intelligence Center, Jack Shanahan, also emphasizes the profound geopolitical ramifications, stating, "AI will fundamentally alter the landscape of warfare and impact national security on a grand scale" [12].

The multi-sectoral applicability of AI is fundamental to many AI companies' goals to create generalizable systems that "can be characterized as 'dual-use'-having both military and civilian applications" [13]. One can even argue that AI systems can be labelled *omni-use*, since many of them can be used across an exceedingly broad range of sectors and domains (perhaps with small amounts of tuning). Horowitz captures this characteristic well, describing AI as "best thought of as an umbrella technology or enabler" [9]. The omni-use nature of AI has its clearest expression in the rapidly expanding ecosystem of enterprises built on foundation models, such as OpenAI's GPT-4 or Meta's open-source LLaMA. These models are intentionally designed to be a base that other entities can fine-tune and build upon for particular uses or applications. The whole point of foundation models is that they are "train once, apply everywhere (with minimal fine-tuning)."

In contrast with other "umbrella" or multi-use technologies, the capabilities and enhancements that AI provides for other technologies have a qualitatively different form. For instance, the military concept of air dominance is only possible because flight technology can be weaponized. However, flight technology does not have a multiplicative effect across all existing technologies in the same way that AI does. Commercial flight technology can enable a missile to fly, but cannot: be used to develop new missile designs; be integrated into targeting operations to process, analyze, and fuse data from multiple sensors; or coordinate and control multiple resources and platforms to strike a target at speed and scale, perhaps even without human control. On the other hand, all of these latter functions can be performed by AI systems, thereby enhancing capabilities across the entire spectrum of commercial or conflict operations [14-16]. The impacts of earlier crossover technologies were largely sector-restricted, while crossover AI—which is to say, *all* AI—has a wholly different level of impact.

When considering real-world military applications of AI by major powers such as Russia, China, and the US, experts point to the ability of AI to be used for multiple purposes such as communications, intelligence gathering, and increasing efficiency across sectors [17]. The capabilities and uses of crossover AI cannot be determined from a single use case or application, unlike earlier crossover technologies (e.g., lasers used to blind adversaries, combustion engines used in tanks, or flight technology to create military aircraft), precisely because such AI can provide novel capabilities across essentially all use cases and applications, if so desired. And of course, the risks of AI technologies are correspondingly broad.

More generally, we can consider AI's impact across what can be labelled "critical technologies." The Australian Strategic Policy Institute (ASPI) Critical Technology Tracker examines the relationships between multiple critical technologies that may impact strategic competition including AI [18]. ASPI's methodology is based on analysis of publicly available academic literature across a range of fields. The Tracker shows that AI can potentially impact military relevant capabilities across many sectors such as advanced aircraft engines, swarming and collaborative robots, and advanced communications [18]. As a result, assessments of critical *non*-AI technology maturity in a particular country must always include consideration of the ways that the country's AI systems could advance those capabilities, but need not to include consideration of other non-AI technologies. Crossover or multi-use technologies are not new, but AI is qualitatively different in this regard from previous non-AI technologies.

## 2.2 The urgency of addressing crossover AI technology

There are a number of concrete examples of civilian AI capabilities that can be, or have been, used for conflict operations with little or no modification. Some well documented examples include Google's AI technology, which was adopted by the DoD to analyze geospatial imagery as part of Project Maven [19, 20] and Microsoft's HoloLens, which has been developed for commercial purposes while simultaneously being tested by the US Army to enhance battlefield awareness of soldiers [21]. A less well known example of civilian AI being used to broadly enhance conflict operations is Palantir's AI Platform for Defense (AIP), which leverages publicly available Large Language Models (LLMs) as part of an automated battlefield assistant. In an eight-minute demonstration video, they show the AIP system marrying a tuned version of the twenty-billion parameter, open-source EleutherAI LLM with real-time battlefield data in order to generate a representation of the environment along with courses of action [22]. That is, the foundation model is used in a military context that is quite different from the original EleutherAI use cases. There is no evidence that the EleutherAI LLM was designed or developed to facilitate operations in conflict settings; nonetheless, the technology has been readily used for exactly those purposes.

More generally, there is currently nothing that can stop a defense contractor, state agency, or non-state actor from using publicly available, open-source foundation or generative AI models to build a system like Palantir's AIP. Obviously, such a system must be linked with (likely) confidential or sensitive data to produce courses of action and recommendations, but such a tool can further leverage publicly available sources of data such as geospatial imagery, images on social media, and a range of other inputs. In fact, one might be able to achieve high-level performance even with purely open-source data, as shown by Ukraine's utilization of the Diia App, a mobile application and online portal that connects Ukrainian civilians to government services [23]. Leveraging Diia, the Ukrainian government has been able to harness the power of an engaged population for wartime situational awareness [24] in a vivid example of the multiplicative effect of crossover AI technology across systems.

One might think that only (relatively) open-source, older, or more limited AI capabilities could be repurposed in these ways, but the speed of technological innovation means that advanced AI capabilities do not remain proprietary or closed for long. Open-source generative AI models, such as Technology Innovation Institute's Falcon, a 180 billion parameter model, can now perform at levels that exceed GPT-3.5 on industry benchmarks, with greater efficiency [25]. That is, almost anyone with access to the Internet can access open-source models that offer the same capabilities that were considered cutting edge only one year ago. Further, there is no indication that this pace of development will decrease; open-source and open-access systems are likely to grow increasingly closer to the cutting-edge frontier. We conjecture that foundation models that support present-day military-grade AI systems are likely to be available to almost anyone via more efficient, open-source models in two to three years.

Going beyond specific use cases, it is important to recognize that states and their militaries have played an active role in the incredible pace of civilian AI innovation, while also strengthening ties with civilian AI companies. Many of the leading AI models are not being developed at academic institutions or government agencies, but at well-funded, commercial enterprises. Both the People's Republic of China (PRC) and the US are actively seeking to facilitate commercial AI innovations that will benefit national defense, often with the declared aim of promoting economic development through integration of civil and military technologies [26]. And the DoD is actively reducing barriers between itself and commercial entities by hiring commercial AI executives [27]. These moves can facilitate the rapid and robust integration of commercial AI technology for conflict operations across fields, again highlighting why crossover AI technology is unlike historical examples. There is good reason for military actors to adopt civilian AI for conflict applications as "most of the biggest global advancements in AI over the past thirty years originated in the commercial sector" [12].

Of course, a civilian company may be largely composed of altruistic individuals who do not intend for their system to be used as a tool for conflict. More generally, even when people recognize that their technology could be used for different purposes, they may question whether they have any corresponding ethical obligations, given that they are not intending to produce a military AI system. As we have shown, civilian AI technology is a crossover technology that can have far-reaching uses beyond the developers' original intentions. However, one might contend that developers do not have ethical obligations for those distant (potential) uses, much as we do not hold a hammer manufacturer ethically responsible for problematic uses of the tool. We thus turn now to consider the moral and ethical responsibilities of actors involved in the development of crossover AI technology.

## 3 MORAL AND ETHICAL OBLIGATIONS WHEN DEVELOPING CROSSOVER AI TECHNOLOGY

AI systems can often be used for a variety of different purposes beyond their original design, but the effort required for such extensions can differ widely across various possibilities. This observation alone implies that civilian developers could potentially have ethical obligations in regards to unintended uses; they might, for example, have an obligation to make it harder (even if impossible is not an option) to extend a system in ways that could cause widespread harms. Civilian non-military entities must be thoughtful about the AI capabilities they are willing to develop, deploy, and integrate, as well as the ethical implications of those choices. Nevertheless, until recently, "technologists seldom deliberated on the societal impacts of their creations, focusing instead on immediate consumer-oriented and profit-driven problem-solving" [12] p. 3. Therefore, this process starts with acknowledgement that AI tools designed for non-military use cases almost certainly will be leveraged by political organizations, government agencies, individuals, and militaries to challenge existing regimes. As argued in the previous section, crossover AI technologies "increase the likelihood that it [AI] will be weaponized by cybercriminals, state-sponsored actors, and lone wolves" [13]. We thus require more than general, high-level observations, and so we ask: **What** are the moral and ethical obligations of actors involved in the development of AI systems that are not intended for military applications?

### 3.1 The moral relevance of reasonably foreseeable outcomes

We contend that stakeholders should be morally responsible for those uses that are *reasonably foreseeable*. The core idea is that an agent's moral responsibility for some action is not necessarily determined by their intentions alone; we must also consider what the agent could reasonably have foreseen to be potential outcomes of their action. As a fantastical example, suppose an evil entity has designed a machine that will kill a random individual if I ever take ten breaths in exactly 50.17 seconds. If I am unaware of the existence of this machine, then I am not ethically liable for a fatality that might occur (even if I am causally responsible for it), as it was not reasonably foreseeable that my act of breathing would lead to such an outcome. More colloquially, the basic idea is that moral agents ought not be held (fully) morally liable for outcomes that they could not reasonably have expected or considered, but should be held (at least partially) morally liable for those that were foreseeable. We emphasize that the core notion here is whether an outcome is reasonably foresee*able*, not whether it was actually foresee*n*. If I throw a rock at someone's head, then I cannot escape moral responsibility for their injury simply by refusing to recognize what would likely occur.

Of course, there are many challenges to precisely stating the relevant foreseeability condition, especially providing a precise characterization of reasonableness that is appropriate for moral agents such as us humans. We are not omniscient beings, and so some potential outcomes are not reasonably foreseeable (e.g., a fatality because of one's breathing rate) while others are (e.g., a rock thrown at someone's head). A full philosophical theory should draw appropriate lines, but those will depend on factors such as the nature of human rationality. In this paper, however, we do not need a precise definition, as the core phenomenon—almost all civilian AI systems could be used to provide or amplify conflict capabilities—is clearly foreseeable for the relevant stakeholders. As discussed in the previous section, the omni-use nature of AI is widely recognized, and AI designers, developers, deployers, procurers, and so forth can thus reasonably foresee that their systems will be used in many ways.

The critical implication is that civilian AI developers *prima facie* bear some measure of responsibility for their AI systems being used for conflict purposes. Moreover, this *prima facie* responsibility holds even if the developers did not want or intend their AI to be used in conflict settings, as such uses are reasonably foreseeable. That being said, we emphasize that this *prima facie* responsibility might be absolved if the moral agent takes affirmative steps to reduce the likelihood of the reasonably foreseeable outcome. We return to this point in Section 4.

More generally, the notion of *reasonably foreseeable outcomes* can be applied by all stakeholders to aid intentional decision making by helping to identify outcomes that have a high potential of occurring. Many ethical theories hold that obligations depend on what is reasonably foreseeable, particularly when we consider evaluation of actions that (causally) create some measure of risk for others (e.g., throwing a rock at someone's head). We propose here to extend the notion of what is reasonably foreseeable to the full lifecycle of AI systems in light of their omni-use nature. As a result, there can be ethical implications of systems that are not being developed for conflict operations, but which are likely to be adopted, adapted, and otherwise integrated into conflict operations – potentially enhancing or supporting an actor's ability to carry out violence.

### 3.2 Reasonably foreseeable consequences of crossover AI technology

We now consider four distinct reasonably foreseeable outcomes of crossover AI technology: 1) AI systems designed for civilian purposes will be used in active conflict, 2) AI systems designed for civilian purposes will be used for conflict support activities, 3) the use of civilian AI tools in conflict operations will impact traditional applications of the law of armed conflict, and 4) AI systems designed for civilian purposes will be used in gray zone conflict.

First, and most obvious, civilian systems that incorporate various AI models can be applied to active conflict. The appeal of incorporating commercial systems in military operations stems from their potential to facilitate both defensive and offensive actions taking place at greater-than-human speed at scale. Complicating the discussion, the line between a commercial AI system designed for "military" versus "civilian" applications is blurry. For example, Franke and Soederstroem highlight how Palantir's AI is being used to automate various processes in the conflict in Ukraine, noting that "the sensor-to-shooter loop – the time from detection of a target to its destruction – has been reduced to just over 30 seconds" [4]. Palantir is a commercial enterprise focused on military clients but, as discussed earlier, their software platforms may rely on fine-tuned versions of open-source AI models. Further, a target can be identified using an unarmed commercial drone, such as a DJI Mavic, running civilian image recognition software, cueing a human who can then rapidly deploy an inexpensive first-person view drone that has been modified to carry out a kinetic strike. For a straightforward, pure software example, we can look to Ukraine's use of Clearview's commercially available facial recognition software for a variety of use cases including identifying deceased soldiers as well as Russian infiltrators [8]. It's also important to note that civilian systems, such as drones that have autonomous capabilities, can be used to carryout physical attacks without attribution, and as Johnson notes, "the ubiquity and rapidly declining unit costs of drones will mean that these capabilities will become increasingly capable, autonomous, and easy to mass-produce." [10]. Similarly, commercially available facial recognition software can be used to identify and target individuals by actors or organizations that are not affiliated with any particular military but are involved in conflict. These examples represent a fraction of the possible use cases for civilian AI for conflict. For all of these reasons, it is reasonably foreseeable that commercial AI systems will be used for physical intrusions or attacks in conflict settings.

Second, it is reasonably foreseeable that civilian AI will be used for conflict support operations. We believe it is important to distinguish conflict support operations from direct conflict activities, though we recognize that these two categories often overlap. The critical point is that it is reasonably foreseeable by civilian AI entities to recognize that their systems may be used by conflict actors even if they have nothing to do with kinetic military operations, including by rogue states and non-state actors. For example, commercial AI systems can be used to sift through data from cell phones, videos, and other sources for details that can be used to support military operations [28]. In fact, US Special Operations Command (SOCOM) has been leveraging off-the-shelf AI for years to fight non-state actors such as ISIS and al-Qaeda [29]. Simultaneously, generative AI can serve as the basis for powerful methods to create misinformation. More specifically, generative AI systems designed purely for entertainment purposes can be used for subversion and propaganda, altering the information environment for civilian populations, government agencies, civil-society organizations, and military actors [28]. All of these developments will push conflict actors to acquire the ability to rapidly parse through and assess amounts of data that exceed human comprehension, securely communicate and effectively utilize valid data, and identify and counter the negative effects of bad data in a dynamic environment. Off-the-shelf AI systems being built by civilian companies will be fundamental to these efforts. Beyond the domains of information and situational awareness, civilian AI systems will be used to aid logistics operations, manage personnel, and enhance many other support activities that must be conducted by any complex organization.

A third reasonably foreseeable outcome is that use of civilian AI tools in conflict operations will dissolve barriers between civilian populations, military organizations, and other actors in ways that impact traditional applications of the law of armed conflict. This could have the effect of making civilian AI systems legitimate military targets. If the most up-to-date information about the operating environment is available through civilian platforms, conflict actors will leverage them. Such systems could potentially become valuable to military operations and therefore a prime target for adversaries. Similarly, the effective utilization of information created through civilian sources can be facilitated by commercially available AI systems that can process and analyze the information being generated. If a civilian uses their mobile phone to take a digital image of a military object, such as an armored personal carrier, and uploads this data to a social media application, it is conceivable that a third-party could use an image recognition system to rapidly identify the image, geolocate the image using open-source software [30], and initiate a military targeting sequence. How various civilian AI systems are used to create or enhance military capabilities will impact how adversaries view them as well as the civilian users or developers.

The law of armed conflict is predicated on being able to distinguish valid military targets. Civilians and civilian structures are generally protected by the principles of necessity, distinction, and proportionality. However, if civilian AI systems are used in conflict with no or minimal changes, then the AI system risks being a valid military target. As commercial AI systems become increasingly important in conflict, "private companies need to prepare for their products, and potentially even their staff, becoming targets in future conflicts" [4].

A fourth reasonably foreseeable outcome is that crossover AI technology will be applied to conflict that takes place between peace and armed conflict. This can be labeled *gray zone conflict,* referring to the ambiguous space "between peaceful competition and armed conflict" [31]. Gray zone activities to weaken or attack an adversary have always existed, but civilian AI systems are particularly useful for gray zone activities, thereby producing new attack vectors that make attribution and detection more challenging for impacted states [32]. For example, the broad range of publicly available generative AI tools make it possible for states, proxies, and non-state actors to produce high-quality disinformation products with little to no in-house technical capabilities. Civilian populations using social media, driven by algorithmic models, can then be leveraged to spread disinformation, undermining legitimate governments that are not involved in open conflict.

One can argue that the principle of proportionality, which protects civilians from intentional targeting under the laws of armed conflict, should also apply outside of armed conflict, thereby providing constraints on the use of AI in gray zone conflict. However, the principle of proportionality becomes difficult to apply if we consider AI systems that are designed for civilian purposes. If commercial AI systems are used by an aggressor, to what extent are the commercial entities that developed the relevant systems responsible? What if a generative AI system developed in Country X is used by Country Y to conduct a disinformation campaign that undermines Country Z's democratic elections or incite riots? If and how a state chooses to retaliate for such an act will be a matter of many variables, but it is likely that crossover AI technology can and will be used to carry out such attacks. One can even argue that it is reasonably foreseeable that the availability of powerful commercial systems makes it possible for a broader range of actors to carry out such attacks (e.g., a non-state actor in Country Y could conduct the disinformation campaign in Country Z). In other words, commercial AI systems will enhance the capabilities of various actors, including non-state actors, allowing them to pose significant gray zone threats.

Neither gray zone activities nor the capabilities of commercial AI tools are confined to information operations or cyber-attacks. For example, with some technical knowledge, an actor could potentially fine-tune an open-source generative AI model to produce formulas for toxic agents that could, if synthesized, be used to physically poison individuals or populations. This possibility was demonstrated in 2022 when researchers from a private drug discovery firm were able to use their own generative AI model to design 40,000 toxic molecules in six hours, demonstrating, "that designing virtual potential toxic molecules is possible without much effort, time or computational resources" [33]. Whether we consider information operations, cyber-attacks, or the misuse of generative AI to create new means and methods of causing harm, it is reasonably foreseeable that crossover AI technology will pose novel gray zone conflict threats.

Given the spectrum of reasonably foreseeable outcomes of crossover AI technology described above, how do we ensure that harmful possibilities are minimized? And if an actor is attempting to misuse or weaponize an AI tool, how do we detect, track, and respond across traditional divides between state and commercial actors or even international borders? These difficult questions must be addressed by all stakeholders involved in the AI system lifecycle, including civilian actors, because the foundation models being developed by commercial enterprises for civilian purposes will be integral aspects of the various systems that will be used to enhance and carry out conflict operations.

## 4 RECOMMENDATIONS TO ADDRESS THREATS POSED BY CROSSOVER AI TECHNOLOGY

AI systems developed for civilian purposes can crossover to enhance conflict operations by a range of actors, and so commercial actors bear a *prima facie* moral responsibility for such uses in both armed conflict and gray zone conflict, including potential harmful outcomes. This moral responsibility can be reduced or eliminated if developers take proactive steps to address the reasonably foreseeable harms. We propose three linked actions that can potentially serve to absolve commercial AI system developers of some moral responsibility for such uses: (a) establishing systematic approaches to multi-perspective capability testing, (b) integrating digital watermarking in model weight matrices, and (c) utilizing monitoring and reporting mechanisms for conflict-related AI applications. We contend that these are all technically feasible, though we are less sanguine about political feasibility.

### 4.1 Establish multi-perspective capability testing

The first concrete action that civilian entities can take to address the crossover potential of AI systems they are developing is to establish and implement *multi-perspective capability testing*. A variety of approaches, often ambiguously referred to as "red teaming" [34], have been adopted from the cybersecurity realm to help identify vulnerabilities, faults, and threats presented by AI systems. However, while "red-teaming serves a very specific role to identify risks and advance AI accountability…it faces substantial limits in mitigating real-world harms and holistically assessing an AI system's safety" [34]. There have been a number of recent proposals to conduct more testing of AI systems (e.g., voluntary commitments by many leading AI companies to the US White House or the 2023 DEF CON Generative Red Team Challenge). And while robust testing is fundamental to identifying national security threats posed by AI systems, including potential uses in conflict settings, the current proposals are incomplete in various ways.

Our proposed *multi-perspective capability testing* can be used to systematically identify ways that AI systems can enhance the real-world capabilities of bad actors through means other than system faults or vulnerabilities. Multi-perspective capability testing is a non-adversarial approach that intentionally aims to diversify system testing, facilitating the identification of risks posed by AI systems that amplify or enhance potentially dangerous capabilities. In other words, multi-perspective capability testing seeks to identify novel threats and opportunities through a systematic and rigorous approach to real-world use of socio-technical systems. Morally responsible entities can use multi-perspective capability testing to determine how their systems enhance the threat capabilities of bad actors, and thereby take steps before deployment to make such uses more difficult or less probable.

In more detail, multi-perspective capability testing is a process in which panels of non-experts use the AI systems, and their performance is compared to that of panels composed of domain experts. For example, one can compare the conflict capabilities of AI systems in the hands of college students versus experienced military commanders. The goal is to determine the marginal benefit of the AI system in terms of increasing the scopes of use and users. Through this method, morally responsible entities can evaluate if a system can amplify a non-trained individual's ability to plan a military-style operation, terrorist attack, or other conflict-related activities. Multi-perspective capability testing can also allow stakeholders to determine the power of their AI systems to support the development and iteration of attack vectors, means, and methods of conflict (including mis- and disinformation campaigns, physical attacks, conflict tactics, and others). As an aside, we note that this approach can also be used to address capability enhancement for contexts and use cases beyond conflict applications (e.g., AI impacts on space operations, biosecurity, etc.).

As proposed, multi-perspective capability testing supports pre-deployment identification of risks posed by a broad spectrum of reasonably foreseeable conflict related uses of commercial AI systems. This form of testing is grounded in the fact that conflict operations are highly iterative. Conflict actors constantly develop new attack vectors, means, and methods that can be effective against existing defenses and counter measures. Commercial entities thus have a moral responsibility to identify ways their systems enhance these abilities and to preemptively address them. Further, iterative approaches to developing conflict-related applications are not confined to state-based military forces. Non-state military actors, including terrorist organizations and criminal entities, may be able to leverage iterative approaches more effectively than law-based and bureaucratically bound actors. For example, homemade bombs developed by terrorist cells can be highly lethal against advanced military armor given key technical know-how and creativity even as new ways of jamming such bombs are deployed as was seen with the use of explosively formed projectile (EFP) bomb technology against US military equipment in Iraq from 2004 to 2011 [35]. Multi-perspective capability testing is a systematic method to preemptively identify the ways AI systems can be used to aide or enhance conflict operations, mitigate the risks presented by these capabilities being leveraged by bad actors, and minimize threats to peace and security while maximizing for positive use cases.

### 4.2 Integrate a collection of digital watermarking practices

The second concrete action that morally responsible developers of commercial AI systems must take is to embed digital watermarks in their systems. Unfortunately, at the current time, digital watermarks in AI outputs can relatively easily be removed or corrupted. We believe that there is a moral responsibility to develop and implement digital watermarking that is not easily removed but will support digital forensics and provenance tracing, a view shared by many organizations (e.g., Executive Orders 13859 and 14110). However, we contend that there are watermarking steps that could be taken now, albeit for models rather than outputs.

To address digital forensics for crossover AI technology, developers should implement a two-part approach to watermarking: 1) embedding digital watermarks within the weight matrices of foundation models, and 2) embedding digital watermarks in outputs that can be used to check provenance given key information. This two-part approach is not a panacea to the challenge of detecting and tracing AI generated material (images, text, etc.) as techniques to strip digital watermarks or spoof detectors are a recognized problem [36, 37]. Nor does this approach facilitate the broader public's ability to identify digital outputs (e.g., fake videos used to spread disinformation). Rather, this approach is focused on making it easier to identify the provenance of an output or system that is being used for conflict applications, counter to the intentions of morally responsible stakeholders.

In more detail, the first part of this recommendation is to embed digital watermarks in system weight matrices. This practice will aid the ability of entities to trace problematic use cases to the foundation models, creating a pathway to address concerns at a root level that may not have been identified prior to system deployment and integration. Such watermarks will be technically challenging to strip out as the watermarks will have a negligible impact on system outputs. At the same time, these signals could be valuable for digital forensics, if it is possible to access the system being used. To the best of our knowledge, no such watermarks exist at the moment, but there are no obvious technical barriers. One might worry that fine-tuning of a large foundational model would eliminate the watermark, but fine-tuning has predictable impacts on the weight matrices, thereby enabling the design of watermarks that persist after fine-tuning (e.g., watermarks based on relative weights rather than absolute magnitudes). This step would enable a certain type of digital forensics: Given a model that was used for problematic purposes, one can trace the model's provenance back to the original developer. This action will allow morally responsible commercial actors to determine in a *post hoc* manner if their foundation model is being used to support the development of systems and use cases that are morally objectionable. It must be emphasized that this practice only supports digital forensics if stakeholders are able to access the model that is being used for conflict applications (which is conceivable for a government-run investigation or military operation against a terrorist organization).

The second part of this recommendation requires developers to embed watermarks in system outputs that are not easily detectable (and therefore not easily stripped out) but which can be identified via a clearinghouse. The clearinghouse will allow developers of open-source or open-access systems to be granted access to a portal whereby they can check problematic material for the digital watermarks they have embedded. Such an approach will aid digital forensics and support the identification of AI systems that are being repurposed by bad actors. At the same time, the use of a restricted clearinghouse can make it harder for bad actors to determine the particular watermarking strategy in some output, and thereby make it harder to know what to remove. This approach is not without challenges. For example, access to the clearinghouse will have to be restricted to prevent bad actors from being able to reverse engineer the watermarking method and then strip out the watermarks. But a morally responsible system developer that participates in the clearinghouse will better be able to identify if and how their model is being used outside of intended use cases.

We emphasize that digital watermarking is not a technical panacea. People can be influenced or impacted in a variety of ways, and remain susceptible to malign influence using AI. Nonetheless, if broadly adopted, the act of digitally watermarking foundation model weight matrices and embedding hidden watermarks in system outputs could enable commercial AI developers to iteratively make it harder to create a problematic use case at a system level, and to enable identification when problematic uses arise.

### 4.3 Create monitoring and reporting mechanisms

Even if an AI system has been thoroughly tested and a method to identify the provenance of the model has been integrated, bad actors will surely attempt to use civilian systems for conflict operations. To address this reality, mechanisms for monitoring and reporting conflict-related activities need to be established and maintained. This action does not require an invasion of user privacy, but it does require intentional choices about where privacy is required and how to address unacceptable levels of risk.

Recent studies have established that publicly available LLMs are vulnerable to jailbreaking [38]. The jailbreaking techniques allow actors to use civilian systems in ways they are not intended, bypassing technical guardrails. A nefarious actor using such jailbreaking techniques can therefore leverage cutting edge models to enhance their ability to carry out conflict operations. Additionally, methods of identifying new risks or use cases and rapidly addressing them must be an integral part of a morally responsible actor's system lifecycle. This means that when vulnerabilities are identified, a transparent plan of action and commitment to address the vulnerabilities must be communicated [34]. As established in Section 3, it is reasonably foreseeable that commercial systems will be applied to conflict operations by a range of actors, therefore commercial enterprises have a moral responsibility to establish transparent policies for system monitoring. Considering the near universal practice of commercializing user data, claims that such monitoring will violate privacy can be rejected as an attempt to bypass responsibility.

However, even if clear and transparent monitoring processes are established, they will be of little value if government bodies do not establish similarly clear and transparent reporting mechanisms. Reporting mechanisms must be government-led with industry and civil society input and monitoring. Information security about actual risks and mitigation efforts will be paramount to reduce the threat posed by crossover AI technology. We envision a clearinghouse that will allow for a pooling of information about vulnerabilities, mitigation efforts, and real-time threat identification best practices. This action will be challenging due to the need to be transparent while also recognizing the sensitivity of the information. However, it is essential that reporting mechanisms be put in place that support the identification and mitigation of new vulnerabilities and threats if we are to address the crossover potential of AI technology and the moral and ethical responsibilities we have as developers and users of powerful systems.

### 4.4 Hypothetical example of the recommendations in practice

Let's now consider how the three linked actions described above will have practical impacts on a commercial AI system. For this simplified hypothetical case, suppose that a publicly available generative AI system is identified by digital activists as being used by a bad actor to support a disinformation campaign that is designed to undermine the credibility of a democratic election as part of a gray zone conflict operation. The digital activists report the use case to a centralized reporting clearinghouse established and maintained by the federal government. One of the AI companies that monitors the reporting clearinghouse identifies their confidential digital watermark in the material that is reported. This allows the commercial AI company to verify that the harmful material is being generated by a fine-tuned version of its foundation model. The company then carries out mitigation actions (i.e., modifies the foundation model, restricts access to the model, etc.) to constrain the ability of the bad actors to use the model for the reported disinformation campaign. Simultaneously, the company's multi-stakeholder capability testing procedures are updated to account for the reported use case, to pre-emptively address such threats in future models. The updated methodology is also reported via the clearinghouse so that other developers can implement the testing methodology and needed guardrails or system use restrictions.

## 5 CONCLUSION

This article represents a first effort to establish a cohesive theoretical approach to the moral responsibilities of commercial entities that are developing AI systems for non-military applications that are likely to be used for conflict applications. Going beyond previous work that discusses the dual-use nature of AI systems, we contribute to the field by showing that actors must address four reasonably foreseeable outcomes. With these reasonably foreseeable outcomes established, we then offer three linked, technically feasible actions that system developers must take to reduce or eliminate their moral responsibility. Future work identifying and categorizing the nuances of reasonably foreseeable outcomes of dual-use AI technology would be helpful. We also recognize that there are a range of challenges to the three linked actions we offer. More research will be required to develop methods and best practices to operationalize the actions. By applying the notion of reasonably foreseeable outcomes, our intent is to offer a philosophical approach that can be used to systematically think about and identify the moral responsibilities associated with developing what is arguably the most powerful, dual-use technology the world has seen.